\documentclass[twocolumn]{aastex62}
\usepackage[utf8]{inputenc}
\usepackage{amsmath}
\usepackage{rotating}
\usepackage{amssymb}
\usepackage{comment}
\graphicspath{{./}{figures/}}

\shorttitle{Disks and Planets}
\shortauthors{Swain et al.}

\begin{document}
\title{Exoplanet System Architecture: Sculpting the Inner Regions}

\correspondingauthor{Mark R. Swain}
\email{mark.r.swain@jpl.nasa.gov}

\author[0009-0001-4487-7299]{Mark R. Swain}
\affil{California Institute of Technology, NASA Jet Propulsion Laboratory}

\author[0000-0001-5966-837X]{Geoffrey Bryden}
\affil{California Institute of Technology, NASA Jet Propulsion Laboratory}

\author[0000-0002-3389-9142]{Jonathan C. Tan}
\affil{Dept. of Space, Earth \& Environment, Chalmers University of Technology, Gothenburg, Sweden}
\affil{Dept. of Astronomy \& Virginia Institute for Theoretical Astronomy, University of Virginia, Charlottesville, VA, USA}

\author[0000-0002-5258-6846]{Eric Gaidos}
\affil{Department of Earth Sciences, University of Hawai’i at M\={a}anoa, Honolulu, Hawai’i 96822 USA}
\affil{Institute for Astrophysics, University of Vienna, 1180 Vienna, Austria}

\author[0000-0002-4891-3517]{George Zhou}
\affil{Centre for Astrophysics, University of Southern Queensland, West Street, Toowoomba, QLD 4350, Australia}

\author[0000-0001-9301-6252]{Caeley V. Pittman}
\affil{Department of Astronomy, Boston University, 725 Commonwealth Avenue, Boston, MA 02215, USA}
\affil{Institute for Astrophysical Research, Boston University, 725 Commonwealth Avenue, Boston, MA 02215, USA}

\author[0000-0002-8828-6386]{Christopher M. Johns-Krull}
\affil{Department of Physics and Astronomy, Rice University, 6100 Main Street, Houston, TX 77005, USA}

\author[0000-0002-3656-6706]{Ann Marie Cody}
\affil{SETI Institute, 339 Bernardo Ave., Suite 200, Mountain View, CA 94043, USA}

\author[0000-0001-7891-8143]{Meredith A. MacGregor}
\affil{Department of Physics and Astronomy, Johns Hopkins University, 3400 N Charles St, Baltimore, MD 21218, USA}

\author[0000-0002-4115-0318]{Laura Venuti}
\affil{SETI Institute, 339 Bernardo Ave., Suite 200, Mountain View, CA 94043, USA}
\affil{Visiting Fellow, School of Physics, UNSW Science, Kensington, NSW 2052, Australia}

\author[0000-0002-4725-7589]{Aayush Gautam}
\affil{Dept. of Astronomy \& Virginia Institute for Theoretical Astronomy, University of Virginia, Charlottesville, VA, USA}

\author[0000-0001-8292-1943]{Neal Turner}
\affil{SETI Institute, 339 Bernardo Ave., Suite 200, Mountain View CA 94043, USA}

\author[0000-0003-3616-6822]{Zhaohuan Zhu}
\affil{Department of Physics and Astronomy, University of Nevada, Las Vegas 4505 S. Maryland Parkway Las Vegas, NV 89154, USA}

\author[0000-0002-7260-5821]{Evgenya Shkolnik}
\affil{School of Earth and Space Exploration, Arizona State University Tempe, AZ 85287, USA}
\affil{Interplanetary Initiative, Arizona State University Tempe, AZ 85287, USA}

\author[0000-0003-1639-510X]{Connor Robinson}
\affil{Division of Physics and Astronomy, Alfred University, 1 Saxon Drive, Alfred, NY 14802, USA}

\author[0000-0002-0267-9833]{Valerie Scott}
\affil{California Institute of Technology, NASA Jet Propulsion Laboratory}

\author[0000-0002-4487-4274]{John Arballo}
\affil{California Institute of Technology, NASA Jet Propulsion Laboratory}

\begin{abstract}
In this study, we seek to improve our understanding of the competing roles of disk-driven and planet-planet dynamical migration in sculpting planetary system architecture in the inner $\lesssim 1.5$ au of protoplanetary disks. Over a range of host star masses, we compare the orbit semimajor axis values of transiting multi-planet and resonant systems to observationally-derived estimates of protoplanetary disk inner truncation radius $R_{i}$, corotation radius $R_{co}$, and dust sublimation radius $R_{dust}$.  We find that disk-driven migration is primarily responsible for setting the inner edge of planetary systems near $R_{co}$ and that subsequent dynamical migration shapes the distribution of planetary semimajor axis values over the range $\approx 20-300$ $R_{\star}$. If multi-planet systems form in a way similar to the resonant chain systems, either a zone of highly efficient planet formation at $\gtrsim 100 R_{\star}$, followed by subsequent disk-driven migration, is implied, or a modified in-situ mechanism operating over a region from $\simeq 15-100 R_{\star}$ and incorporating disk-driven migration is needed. There are indications that after disk dispersal, dynamical migration causes a subset of planets to migrate to locations inside $R_{co}$. 
\end{abstract}

\keywords{exoplanet, atmospheres, exoplanets, formation, protoplanetary disks, migration}

\section{Introduction}
\label{sec:intro}

ALMA observations have revealed that the conversion of protoplanetary disk solids in planets is an efficient process \citep{najita2014} and a potential link between structured disks and the formation of large planets \citep{vandermarel2021}. High resolution ALMA observations have allowed the probing of dust distributions in protoplanetary disks down to 0.6 au ($\approx 130 \: R_{\odot}$) scales and have revealed that compact disks are common \citep{guerra-alvarado2025}. However, ALMA observations do not resolve the scales where the bulk of the currently known transiting planets are located. In this study, we concentrate on the inner regions of protoplanetary disks. 

One of the major questions for planet formation scenarios is whether the individual planets in compact, multi-planet systems formed in-situ \citep[e.g.,][]{2012ApJ...751..158H,2013MNRAS.431.3444C,chatterjee2014}, that is at locations close to the present semimajor axis values, or whether they formed at substantially larger semimajor axis values and then underwent significant inward migration \citep[e.g.,][]{2012ARA&A..50..211K,2015A&A...582A.112B,2021A&A...656A..69E}, or through a combination of these processes \citep[e.g.,][]{burn2024}. The difference between these two formation mechanisms has the potential to lead to profoundly different planetary bulk composition outcomes because of the thermally induced, radial composition gradient in the protoplanetary disk \citep{oberg2011,oberg2016,2022MNRAS.517.2285C,bergin2023}. Multi-planet systems also exhibit radius correlations \citep{weiss2018}, sometimes termed the ``peas in a pod" effect, that is astrophysical in origin \citep{weiss2020} and that must reflect some aspect of the formation process. 

To gain a better insight into the roles of planet formation, migration, and disk properties in determining planetary system architecture, we undertake a comparison of planet semimajor axis values to observed properties of protoplanetary disks. Our study focuses primarily on multi-planet systems because they provide more constraints, but we consider single planet systems as well. In what follows, we describe our methods, report the results, and discuss the implications.

\section{Methods}\label{sec:methods}

The approach taken in this work is to compare the observed semimajor axis values, $a$, of planets in transiting multi-planet systems to observationally determined estimates of two key protoplanetary disk properties: the disk inner truncation radius, $R_{i}$, and the disk corotation radius, $R_{co}$. We also use the observationally estimated luminosity to estimate the dust sublimation radius, $R_{dust}$. Although the main focus is on the multi-planet systems, we extend the approach to include single planet systems. 

\subsection{Planets}

\begin{figure}[t]
\hspace{-0.25in} \\
\centering
\includegraphics[scale=1.0, angle=0]{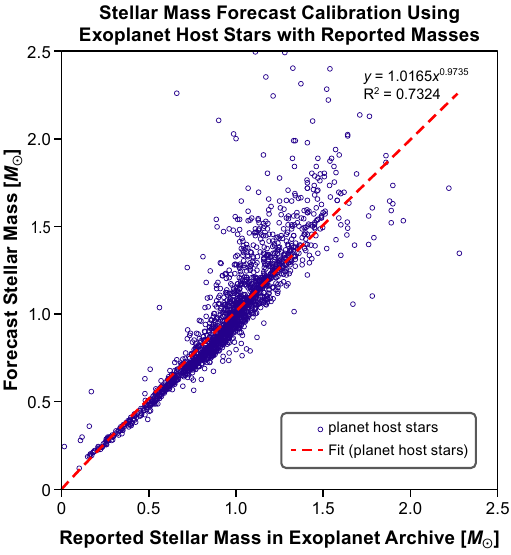}\\
\caption{A comparison of the reported $M_{\star}$ to the forecast value of $M_{\star}$ for exoplanet host stars. The forecast relation (see text) is needed because less half the multiplanet host stars used in our study have reported masses.} 
\label{forecast_test}
\end{figure}

The values of planet semimajor axis and host stellar mass were taken from the NASA Exoplanet Archive using the following filters: archive default values (default\_flag=1), single host star (sy\_snum=1), transiting planet (trans\_flat=1), published and confirmed solution (soltype `Published Confirmed'), planet not flagged as controversial (pl\_controv\_flag=0), semimajor axis value not null (pl\_orbsmaxstr not null), star radius not null (st\_radstr not null). We separated the planets into multi-planet and single planet system lists resulting in sample with 1146 planets in 433 multi-planet systems and 862 single planet systems. More than half of the planets in multi-planet systems listed in the archive did not have a reported stellar mass. Using the Exoplanet archive sample of exoplanet host stars with reported mass and radius values, with a broken power law modeling approach similar to \cite{eker2018}, we identified a forecast relation, $R_{\star} = M_{\star}$ for $M_{\star}\leq M_{\odot}$ and $R_{\star}=M^{0.75}$ for $M_{\star}>M_{\odot}$ ( $M_{\star}$ and $R_{\star}$ are normalized to solar values). We validated the forecast relationship by comparing the reported stellar masses to the forecasted stellar mass (see Fig.~\ref{forecast_test}) and, for uniformity, applied it to the entire sample. As discussed in \cite{eker2015,eker2018}, the main sequence range of $R_{\star}$ values for stars with a given $M_{\star}$ becomes dominated by age for $M_{\star}\gtrapprox0.8 M_{\odot}$ which shows up as increased scatter around the trend in Fig.~\ref{forecast_test}, but this does not impact the main findings of our analysis.

For the single planet systems, we used the Archive default values without modification. For multi-planet resonant systems, a more nuanced approach is required because the Archive default parameter solutions were not available for all known multi-planet systems and the Archive default parameters are not guaranteed to be Keplerian \citep{decocq2026}, which can lead to differing values of $P^{2}/a^{3}$, where $P$ is the planet period, within the same system.  For the multi-planet systems, we augmented the Archive default parameters by Keplerian corrections to the semimajor axis values to enforce a uniform host star mass, $M_{\star}$, for all planets in the system. We also found that some multi-planet systems that have been studied and identified as resonant chains did not have semimajor axis entries in the Archive; for these cases, we computed the semimajor axis values based on the planet orbital period and host star mass. Resonant and near-resonant multi-planet systems (Table 1) are of special interest \citep{siegel2021} because they can be assembled by disk-driven migration \citep{ward1997} and are believed to be the least dynamically disturbed systems. We conducted a literature search and identified 24 systems, containing 100 planets, that have previously been reported as containing planets in resonant or near-resonant configurations (see Table~\ref{tab:resonant_1}). Resonant and near-resonant systems may not have all planets in a resonant configuration and five planets (Kepler-80 f,g, Kepler-90 h, TIO-178 b, TOI-2076 e) in our sample resonant and near-resonant systems are not in a resonant or near-resonant configuration with a neighboring planet and we exclude these planets from resonance-specific analysis.  

\subsection{Disks}

Observational determinations of $R_{i}$, $R_{co}$, and the host star mass and radius were taken from the \cite{pittman2025b} ODYSSEUS survey which includes 47 classical T Tauri systems with measured pre-main sequence (PMS) stellar masses, $M_{PMS}$, hereafter referred to as the ``disk sample''. Disk properties are evolving in time as the star contracts and spins up and exchanges angular momentum with the disk; our disk sample is a heterogeneous snapshot that may not correspond to the epoch of planet formation. However, we can use this disk sample to investigate if there are properties of T Tauri system disks that influence planetary system architecture by looking for correlations between disk properties and planet semimajor axis values. 

Observationally motivated estimates of the dust sublimation radius were made using the ODYSSEUS sample reported luminosity and the relation  
$R_{dust} = 0.07 \, (L_{star}/L_{\odot})^{1/2}$ au, which corresponds to 1500 K dust irradiated by both the star and by the inner edge of the disk itself \citep[so-called dust backwarming;][]{dullemond2010}.  We explored the range of ages for the T Tauri systems in our sample by comparing the disk sample PMS stellar radii to theoretical PMS stellar model isochrones \citep{bressan2012,baraffe2015} and found that that our disk samples $R_{PMS}$ are generally consistent with an age range of $0.5-10$ Myr (see top panel in Figure~\ref{factor}). The luminosity spread also influenced dispersion in early accretion during protostar formation/evolution \citep{baraffe2009,baraffe2012}. The ODYSSEUS survey values for $R_{i}$ and $R_{co}$ are reported in units of the PMS host star radius $R_{PMS}$. 

\subsection{Planet-Disk Properties Comparison}

\begin{figure}[t]
\hspace{-0.25in} \\
\centering
\includegraphics[scale=1.0, angle=0]{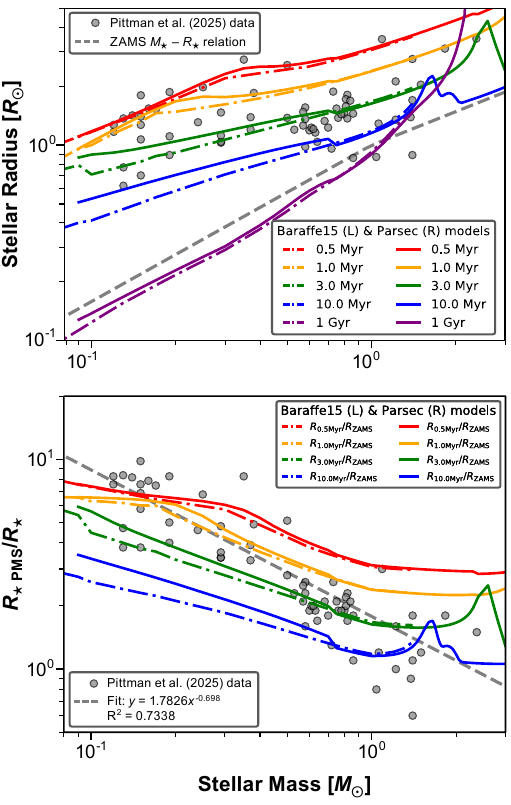}\\
\caption{Top: PMS stellar models \citep{baraffe2015,bressan2012} compared to the disk-hosting PMS star sample used in our analysis. Most of the PMS stars lie in the range of $0.5-10$ Myr. Bottom: The empirically-derived relationship between stellar mass and the T Tauri phase $R_{PMS}/R_{\star}$ stellar radius compared to theoretical model forecasts; the comparison suggests that the fit to the PMS star sample is similar to the relationship between the stellar radius at 1 Myr and the main sequence stellar radius. } 
\label{factor}
\end{figure}

We identify the stellar radius as a natural length scale for examining disk properties and planet orbits over a range of stellar masses. However, we are comparing objects at very different ages and so the question of which stellar radii to use becomes relevant. The majority of the exoplanet systems used in this study have main sequence (MS) host stars whereas our disk sample consists of PMS host stars. We can define two reference frames for representing planet semimajor axis and disk properties $[R_{i},R_{co},R_{dust}]$: one frame is based on the main sequence stellar radii, $R_{\star}$, and the other frame is based on the PMS stellar radii, $R_{PMS}$, consistent with the T Tauri phase objects in the \cite{pittman2025,pittman2025b} survey. Because we have mass estimates for the disk host stars, the transformation to represent the results in the main sequence reference frame is well posed for individual disk targets \citep{pecaut2013}; in contrast, the transformation to represent the results in the PMS reference frame is not well posed for individual exoplanet host stars. However, by using our disk host star sample, we can identify a function, $R_{PMS}=1.5969\times R_{\star}^{-0.812}$ ($R^{2}=0.7437)$, for transforming the exoplanet systems into the PMS host star reference frame (see lower panel in Figure~\ref{factor}). Using this relationship, the 1-$\sigma$ fractional uncertainty on the forecasted PMS radius of an individual main sequence star is 33\%, which for purposes of estimating a sample mean, averages down by square root of the number of exoplanet systems in our analysis. An important caveat is that the function we derived for transforming stellar radii applies to the relationship between the T Tauri phase and the main sequence phase, which can be an important consideration for lower-mass stars with relatively long contraction times to reach the main sequence. 

Depending on the question being investigated, there are advantages for representing the results in either the MS, or the PMS, host star frame. Some parameter trends are better defined visually in one or the other frames and we provide an initial results summary figure based on both the main sequence and the PMS host star frame. In the subsequent analysis, we start in the PMS host star frame and then transition into the MS host star frame.

While the overwhelming majority of planets in our sample orbit MS stars, there is the potential that young planets in our sample have host stars with reported radii that are larger than the their MS radii. To investigate this, we assembled a list of young planets with ages ranging from 3.3 Myr to 290 Myr, which were a combination of single planet and multi-planet systems. To test if the young planet host stars', stellar radii were significantly different from their main sequence values, we used the reported radii, and  the forecast relationship for $M_{\star}$ from Section 2.1, compared the radius-based stellar mass forecast to the reported stellar mass, and found  five young exoplanet host stars where the stellar mass forecast differed from the reported stellar mass by more than 25$\%$; the reported ages for these stars is $\leq23$ Myr. For these five stars (AU Mic, IRAS 04125+2902, K2-33, TOI-1227, \& TIC 88785435), we updated the stellar radius values to reflect the main sequence stellar radius scaling with stellar mass.

\begin{table*}
    \centering
    \caption{Transiting planets in resonant and near-resonant systems used in this study.}
    \begin{tabular}{|l|c|c|c|}
        \hline
        System & $R_{p}$ [$R_{\oplus}$] & Resonant Ref & Parameters Ref \\
        \hline
        AU Mic & $4.0-R_{p}-2.5$ & \cite{dai2024} & \cite{wittrock2023}\\
         HD 63433 & $1.1-2.1\sim2.3$ & \cite{dai2024} & \cite{capistrant2024} \\
        HD 109833 & $2.9-2.6$ &\cite{dai2024} & \cite{wood2023} \\ 
        HD 110067 & $2.2-2.4-2.9-1.9-2.6-2.6$ & \cite{luque2023}&--\\   
        K2-32 & $1.2-5.3-3.1-3.5$ & \cite{heller2019} &\cite{lillo-box2020} \\
        K2-138 & $1.5-2.3-2.4-3.4-2.9-3.0$ & \cite{cerioni2023} & \cite{christiansen2018} \\
        Kepler-51 & $7.1-9.0-9.7$ & \cite{masuda2014} &--\\
        Kepler-60 & $1.7-1.9-2.0$ & \cite{siegel2021} & \cite{jontof-hunter2016}\\
        Kepler-80 & 1.2 $1.5-1.6-2.7-2.7$  1.1 & \cite{macdonald2016}&--\\
        Kepler-90 & $1.3\sim1.5\sim1.3\sim2.9\sim2.6$ $2.8-7.7 \sim 11.3$ & \cite{contreras2018}& \cite{weiss2024} \\
        Kepler-223 & $3.0-3.4-5.2-4.6$ &\cite{mills2016}&--\\
        Kepler-226 & $1.6\sim2.3-1.2$ & \cite{quinn2023} & \cite{rowe2014}  \\
        Kepler-254 & $3.9-2.2-2.5$ & \cite{quinn2023} & \cite{rowe2014} \\
        Kepler-289 & $2.5\sim3.0\sim11.2$ & \cite{dai2023} & \cite{schmitt2014} \\
        Kepler-363 & $1.2-1.7-2.1$ & \cite{quinn2023} & \cite{rowe2014}  \\
        Kepler-1542 & $0.7\sim0.8\sim0.8\sim0.9$ & \cite{quinn2023} & \cite{morton2016} \\
        TIC 434398831 & $3.5-5.6$ & \cite{dai2024} & \cite{vach2025}\\
        TRAPPIST-1 & $1.1-1.1-0.8-0.9-1.1-1.1-0.8$  & \cite{luger2017} &\cite{gillon2016} \\
        TOI-178 & 1.2 $1.8-2.7-2.3-2.4-2.9$ &\cite{leleu2021}&--\\
        TOI-270 & $1.3-2.3-2.0$ & \cite{christodoulou2025} & \cite{gunther2019} \\ 
        TOI-1136 & $1.9-2.8-4.6-2.6-3.9-2.5$ & \cite{dai2023}& --\\
        TOI-2076 & 1.4 $2.8-3.7-3.4$ & \cite{dai2024} & \cite{polanski2024} \\
        TOI-6109 & $4.9-4.8$ & \cite{dai2024} & \cite{dattilo2025} \\
        V 1298 Tau & $5.2-6.4\sim9.7-8.4$ & \cite{dai2024} & \cite{suarez2022}\\
        \hline
        \multicolumn{4}{|l|}{Note: $-$ indicates resonant config., $\sim$ indicates near-resonant config, a space indicates nonresonant config.} \\
        \hline
    \end{tabular}
\end{table*}
\label{tab:resonant_1}

\subsection{Results}

\begin{figure*}[t]
\hspace{-0.25in} \\
\centering
\includegraphics[scale=0.75,angle=0]{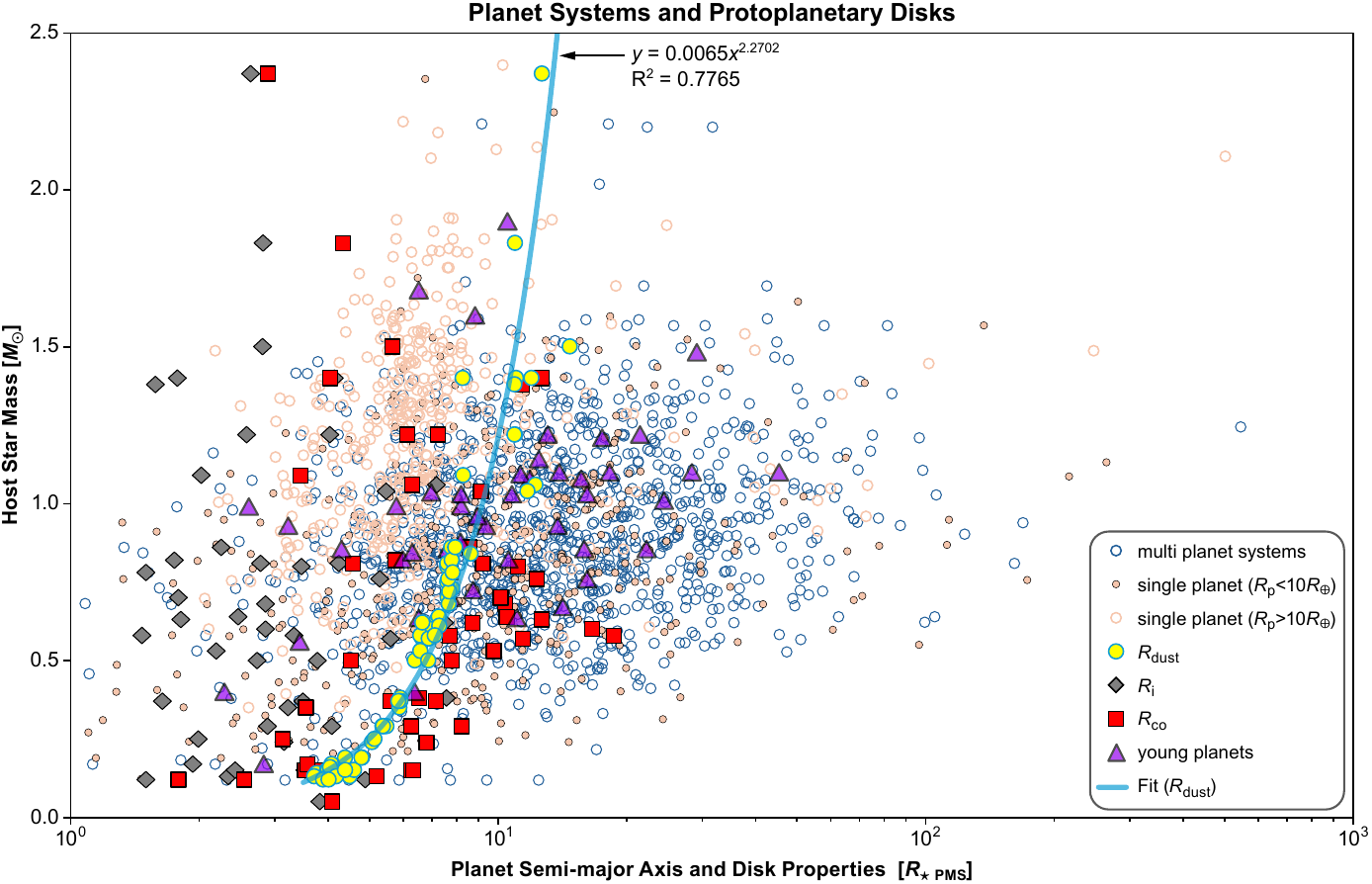}
\caption{Disk properties [$R_{i}$,$R_{co}$,$R_{dust}$] and exoplanet semimajor axis values are plotted in units of the PMS host stellar radius as a function of host star mass. The plot shows planets are less common inside the disk corotation radius suggesting that planetary inward migration is frequently halted in the vicinity of the corotation radius. Some planets have semimajor axis values $< R_{dust}$ the dust sublimation radius, implying that planet migration must have occurred to populate the semimajor axis $<R_{dust}$ region. $R_{i}$ \& $R_{co}$ (gray diamonds and red squares respectively) are from the \cite{pittman2025} survey.}
\label{results}
\end{figure*}

As previously discussed, for the comparison of disk properties to exoplanet system properties, we find it useful to represent the semimajor axis values of planets in units of the host star in either MS or PMS units. When working in units of the host star radius (either MS or PMS), the previously reported dependence of the semimajor axis of the closest planets on stellar mass \citep{sun2025} vanishes; this supports the concept that the stellar radius is a natural length scale for comparing disk properties and exoplanet orbit properties. The comparison of planetary semimajor axis values to disk properties is shown in Figure~\ref{results}.

\section{Discussion}

Major themes connected to Figure~\ref{results} that we explore in this work are protostellar disk properties, disk-driven migration, dynamical sculpting of planetary systems, planet formation regions, and differences between single planet and multi-planet systems. We discuss these themes below.

\subsection{Disk-driven Migration}\label{migration}

\begin{figure}[t]
\hspace{-0.25in} \\
\centering
\includegraphics[scale=1.0, angle=0]{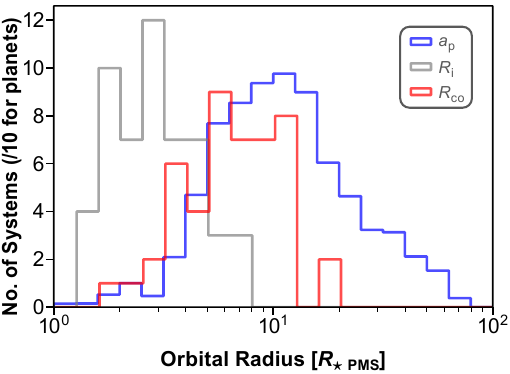}\\
\caption{An observational-bias corrected histogram comparing the location of the semimajor axis values of the innermost planets in multi-planet systems to the location of the corotation radius and the inner truncation radius. The location of the innermost planet in multi-planet systems (blue) is highly correlated with the corotation radius (red), suggesting that disk-driven migration is halted in the region of the corotation radius rather than in the region of the inner truncation radius (gray).} 
\label{histogram}
\end{figure}

One of the most striking features in Figure~\ref{results} is the apparent decrease in the number of planets inside of $R_{co}$. To quantify this visual impression, we corrected the transiting planet observational bias by weighting the innermost planet in each system by the semimajor axis in units of $R_{\star}$ of the outermost transiting planet in the multi-planet system, because the outermost transiting planet defines the system inclination and thus the probability observing a given system; we then constructed a histogram of the semimajor axis values of the innermost planet of multi-planet systems and compared this to the histograms of $R_{i}$ and $R_{co}$. For the range of stellar masses for which we have disk properties, the results suggest that $R_{co}$ has a key role in establishing the location of the innermost planet and thus in planetary system architecture (see Figure~\ref{histogram}) as evidenced by the alignment of the semimajor axis distribution of the innermost planet in multi-planet systems with the distribution of $R_{co}$. We quantify this by comparing the inner edge of the distributions in Figure \ref{histogram} (orbital radii $< 10 R_{\star}$ via a two-sample Kolmogorov-Smirnov (KS) test, finding a 56\% probability that $a_{p}$ and $R_{co}$ are pulled from the same distribution.
This is in strong contrast to the $R_i$ distribution which has a $10^{-18}$ probability of matching $a_{p}$. Our findings for the location of the inner most planet in multi-planet systems are consistent with \cite{mulders2018} and suggest that physical processes in the vicinity of the $R_{co}$ are important for halting inward migration.

While the region dominated by the disk property $R_{i}$ $\approx1.5-3.0 R_{PMS}$, appears as largely unpopulated by planets in Figure~\ref{histogram}, careful inspection of Figure~\ref{results} reveals that there are a few planets in this region---a point that we will revisit later. The most extreme cases are six planets orbiting stars $\leq0.7 M_{\odot}$, that have semimajor axis values of $\leq$ 1.3 $R_{PMS}$. It is worth noting that $R_{i}$ can potentially fluctuate on short timescales \citep{pittman2025} as its location is determined by a combination of the instantaneous accretion rate and the PMS magnetic field strength. The accretion rate can fluctuate on the time scale of hours whereas the magnetic field is expected to fluctuate on time scales of order months.

The substantial decrease in planet occurrence in the vicinity of the inner edge of the $R_{co}$ distribution strongly suggests a scenario in which planets frequently undergo disk-driven migration that is subsequently halted near $R_{co}$ (see Figure~\ref{histogram}); this type of scenario has been explored theoretically \citep{raymond2014,lee2017,izidoro2021,izidoro2022}. If disk-driven planet migration is being halted at or near $R_{co}$, how is this being accomplished? Theoretical work identifies a potential mechanism for halting migration in the form of modulating, or changing the sign of, corotation torques \citep[e.g.,][]{paardekooper2009,paardekooper2011,hellary2012,bitsch2015,yu2023}; some work suggests that the $R_{co}$ region is where the stellar magnetic field coupling to the disk starts to become significant \citep{batygin2013}. 

\begin{figure}[t]
\hspace{-0.25in} \\
\centering
\includegraphics[width=0.45\textwidth]{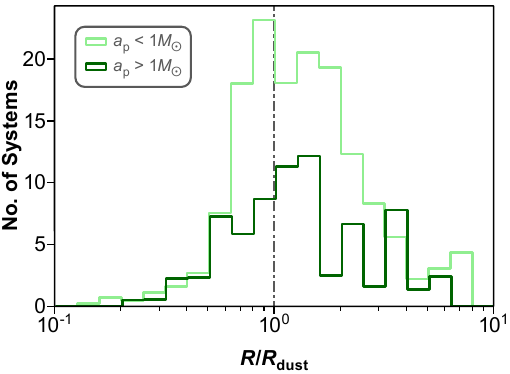}\\
\caption{The observation-bias corrected occurrence rate for the combined sample of innermost planet in multi-planet systems and single planets correlates well with the $R_{dust}$ trend model. There are some signs that for stars with masses $>1~M_{\odot}$, planets tend to move through the dust sublimation region but this may be due to the high density of hot-Jupiter type planets in the vicinity of $R_{co}$ for stars in this mass range.} 
\label{histo_dust}
\end{figure}

Another possible reason for disk-driven migration to be halted near $R_{co}$ is that it is approximately co-located with the dust sublimation radius. While $R_{dust}$ is model dependent, the prescription used in this work results in $R_{dust} \sim R_{co}$ (see Figure~\ref{histo_dust}). A study by \cite{benitez-llambay2018} suggests that under the right conditions, dust torque can halt planet migration and perhaps these conditions are met near the dust sublimation radius. A close inspection of Figure~\ref{results} suggests that single planets with $Rp>10\:R_{\oplus}$ can be located inside the dust sublimation radius, consistent with the findings of \cite{mendigutia2024}.

A third possibility is that $R_{i}$ was located closer to $R_{co}$ during the early part of the T Tauri phase \citep{gaidos2025}; if the planetary migration occurred sufficiently early in the system's history, halting planets at $R_{i}$ would appear, if the same systems were observed later in the T Tauri phase, as planets halting at $R_{co}$. 

Thus, there are several potential scenarios that could result in the system architectures we observe, and these require further investigation. The physical mechanism by which planet migration is halted cannot be determined solely by correlation analysis from the data used in this study. All we can say with certainty is that the observationally determined $R_{co}$ marks a significant transition in the exoplanet semimajor axis distribution.

The concentration of single giant planets, defined here as $R_{p}>10R_{\oplus}$, along $R_{co}$ (Figure~\ref{results}) is interesting for two reasons: First, it suggests that if giant planets underwent high eccentricity migration, this migration must have occurred while the protoplanetary disk was in place. Second, the majority of the multi-planet system planets have $R_{p}<10 R_{\oplus}$ and thus multi-planet systems do not provide a stringent test of a scenario in which giant planet migration is halted at $R_{co}$. The potential halting of the inward migration of planets of a wide range of sizes (and masses) suggests that the the corotation region of the protoplanetary disk could be strongly diffusive and prevent the corotation torque from saturating \citep{paardekooper2011}. There is also the possibility that magnetospheric rebound pushes planets back towards $R_{co}$ and removes planets from the region interior to $R_{co}$ \citep{beibei2017}.

\begin{figure}[t]
\hspace{-0.25in} \\
\centering
\includegraphics[scale=1.0, angle=0]{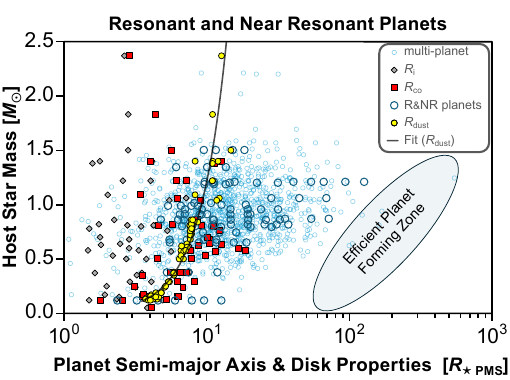}\\
\caption{Resonant and near-resonant exoplanets with semimajor axis values in units of the pre-main sequence stellar radius. The innermost planet in resonant and near-resonant systems is frequently inside the estimated dust sublimation radius, implying that at least some migration must have occurred and was subsequently halted at $R_{co}$.  If resonant chain systems and multi-planet systems are formed by convergent migration, the zone of efficient planet formation must lie further out in the protostellar disk.} 
\label{resonant}
\end{figure}

As is readily apparent in Figure~\ref{results}, planets from both single and multi-planet systems have semimajor axis values that are inside $R_{dust}$, which would require migration. In principle, the migration could be disk-driven or dynamical (by which we mean planet-interaction-driven). One way to attempt to probe pure disk-driven migration is to consider systems that are in resonant or near-resonant chains as these are believed to be assembled by a disk-driven, convergent migration process \citep{cresswell2008,wong2024} and to have undergone little or no additional dynamical migration, which would disrupt the resonant chain. Multi-planet resonant chains are relatively rare but several have been identified (see Table 1), which we plot together with the disk properties in Figure~\ref{resonant}.

The majority of the resonant chain systems have one or more inner planets that are inside the estimated dust sublimation radius, where planets are unlikely to have been able to form in-situ due to the absence of solid material. Thus the presence of resonant chain planets inside the dust sublimation radius indicates that these planets must have formed at semimajor axis values larger than $R_{dust}$ and then undergone subsequent disk-driven migration. A caveat here is that there are alternative estimates of $R_{dust}$, i.e., if the temperature is set by local viscous heating, as invoked in some models \citep[e.g.,][]{chatterjee2014}; in this particular case, the location of the innermost planet would be at $T\approx1200\:$K and $R_{dust}$ would be slightly interior to this. Another noteworthy aspect of the resonant systems is that we find that none of the innermost planets have semimajor axis values that are substantially inside of the $R_{co}$ distribution; this independent line of evidence suggests that the disk-driven migration process weakens or is halted in the vicinity of $R_{co}$. 

The comparison of disk and planet semimajor axis values would benefit from additional observational studies of disk properties by filling in the mass ranges that need better sampling and by improved sampling of the range of values taken on by $R_{i}$ and $R_{co}$ as a function of stellar mass. Furthermore, disk properties cannot be assumed to be constant due to changes in the accretion rate, stellar magnetic field strength, and stellar angular momentum. This potential for variations strengthens the argument for larger surveys to gain a better understanding of the range of values that $R_{i}$ and $R_{co}$ can take on as a function of stellar properties.

\subsection{Dynamical Migration}\label{dynamic}

In addition to looking for signatures of disk-driven migration, we also searched for evidence of migration induced by planet-planet gravitational interactions, which we refer to as dynamical migration. Although disk-driven migration may be the dominant process while the disk is present, dynamical migration can occur on longer timescales, allowing us to investigate the potential for dynamical migration with young planets and by comparing the sample of multi-planet systems to resonant planets. 

Relatively few young transiting planets are known. By considering both single and multi-planet systems, we were able to assemble a list of 45 young planets with host star ages of less than 300 Myr and with a median age for the sample of 200 Myr. The young planet sample contains 25 planets from multi-planet systems and 20 single planets; six of the planets in the sample have $R_{p}>10R_{\oplus}$. All of the planets with ages $<300\:Myr$ from the \cite{dai2024} sample are include. All of the multi-planet systems, with an age of $<50\:Myr$ in our sample of young planets, are resonant systems. In creating the young planet sample, Kepler-1663 was omitted because the age was listed as zero, and Kepler-411 were omitted because its age estimate is based on gyrochronology.

\begin{figure}[t]
\hspace{-0.25in} \\
\centering
\includegraphics[width=0.45\textwidth]{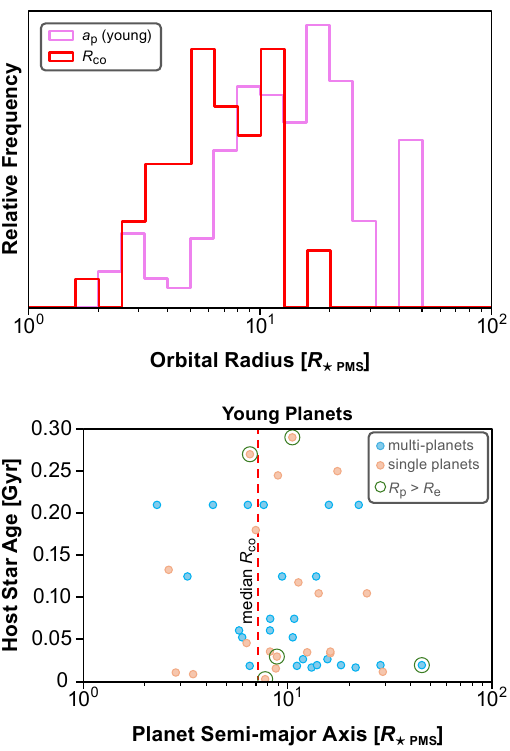}\\
\caption{{\it (a) Top:} The observation-bias corrected young planet semimajor axis distribution is consistent with disk-driven migration halting at or near the disk corotation radius. {\it (b) Bottom:} The young planet distribution of semimajor axis vs age suggests that a some of them may have undergone post-disk dynamical migration on time scales of $\gtrsim 100$ Myr.} 
\label{young}
\end{figure}

We compared the observation-bias corrected (scaling by $a/R_{\star}$) semimajor axis distribution of young planets to distribution for $R_{co}$ and found that young planet distribution is consistent with migration being halted near the disk corotation radius (see Figure~\ref{young}a). We also plotted young planets semimajor axis vs age in the PMS frame and we find a potential deficit of young planets with the smallest semimajor axis values (see Figure~\ref{young}b). For systems that have ages $>100\:Myr$ there are four planets well inside $R_{co}$ whereas for systems with ages $<100\:Myr$ there is only one. While the sample size is very small, the pattern suggests that planets first undergo disk migration, which halts in the $R_{co}$ region, and then a smaller number undergo dynamical migration that takes place on timescales of $\sim 100$ Myr (see Figure~\ref{young}). Due to the small sample size, and the potential for an observational bias due to the difficulty of detecting planets around very young stars, the suggestion from the data is intriguing but cannot be considered as strong evidence, and we note this is a topic to be revisited when the sample of young planets has been enlarged.

\begin{figure*}[t]
\hspace{-0.25in} \\
\centering
\includegraphics[scale=0.75, angle=0]{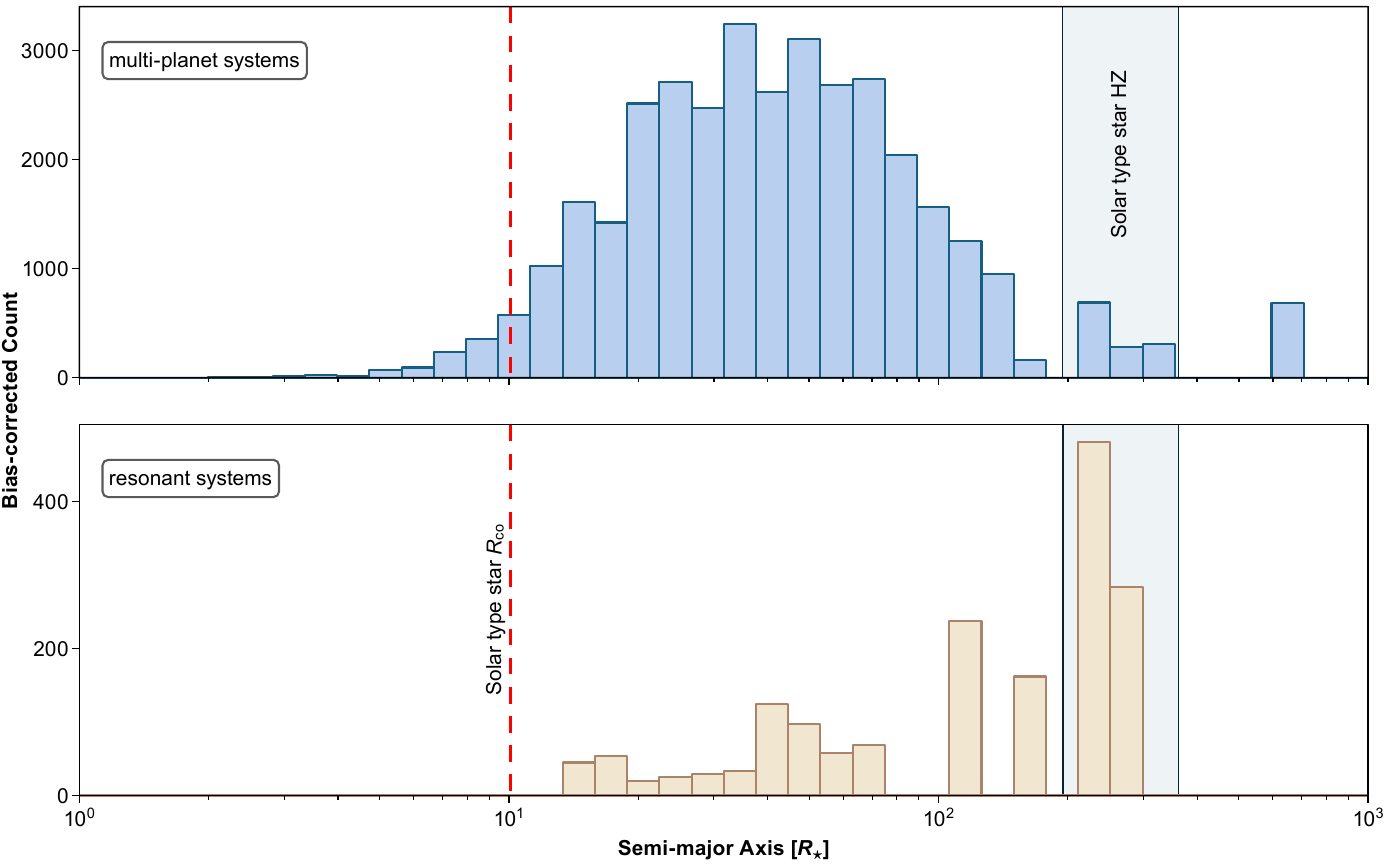}\\
\caption{The observation bias-corrected planet occurrence as a function of semi-major axis for multi-planet systems and resonant and near-resonant systems. If multi-planet systems are initially assembled as resonant systems \citep{dai2024}, then the majority of multi-planet systems must undergo subsequent evolution that profoundly changes the system architecture.} 
\label{sma_histo}
\end{figure*}

To search for evidence of dynamical migration using the multi-planet systems, we constructed observational-bias-corrected histograms for planet occurrence as a function of semimajor axis value. The histograms, shown in Figure~\ref{sma_histo}, were made for planets in multi-planet systems and the resonant system planets and we note the substantial difference in histogram shape. The observation bias corrected semimajor axis histogram  peaks at $\approx 40 R_{\star}$ for the multi-planet systems and at $\approx 200 R_{\star}$ resonant systems; comparing these the non-bias-corrected samples with a KS test yields a score of 0.3462, corresponding to a p-value of 0.0064, implying these are different distributions. 

Recent work suggests that close packed multi-planet systems are initially assembled as resonant chains and that subsequent dynamical evolution produces non-resonant configurations \citep{dai2024} and our findings strongly support this picture. After applying an observation bias correction of $a/R_{\star}$, we find the distribution of planet occurrence as a function of semimajor axis is completely different for resonant and multi-planet systems and is broadly consistent with previous work interpreting period ratios in Kepler multi-planet systems \citep{steffen2015}. Figure~\ref{sma_histo} also shows that none of the resonant system inner planets lie inside of $R_{co}$ while a small fraction of the multi-planet population does. This suggests that some of the inner planets in multi-planet systems undergo dynamical evolution to semimajor axis values $<R_{co}$. If we interpret the resonant systems as representing the disk-driven end state of exoplanet system assembly inside of $\lesssim 1.5$ au, then we would conclude that dynamical evolution:
\begin{itemize}
    \item profoundly sculpts system architecture and concentrates planet semimajor axis values in the $20-80$ $R_{\star}$ region,
    \item produces a relatively small number of planets with semimajor axis values $<R_{co}$ such that the initial imprint of the disk $<R_{co}$ on determining the distribution of the inner planets is retained,
\end{itemize}
Dynamical evolution appears to remove a large fraction of the resonant planets with semimajor axis values $\geq 100$ $R_{\star}$ and redistribute those planets while simultaneously disrupting planetary mean-motion resonances interior to $\sim 100$ au. The role of planet-planet scattering in creating tightly packed exoplanet systems has been previously studied \citep[cf][]{raymond2009} as well as the possibility of planet-planet scattering increasing the range of semimajor axis values, and mutual inclinations, in multi-planet systems \citep{izidoro2017}; our findings are consistent with both scenarios.

However, a potential alternative scenario for the large difference in the planet semimajor axis distributions for multi-planet systems and the resonant systems is one in which the formation process for the resonant systems is fundamentally different from that forming the majority multi-planet systems. Whether such a scenario can be made consistent with the results from \cite{dai2024} is beyond the scope of this paper, and we identify this as an area for future study.

We also note that the semimajor axis distribution of resonant planets indicates that the habitable zones of solar-type stars have, during the period of initial system assembly, a relatively high planet occurrence rate. This will be discussed further in Section 3.5.

We conclude that  dynamical evolution can explain planets with semimajor axis values $<R_{co}$. Previous work \citep{lee2017} has explored the possibility that tidal migration could be playing a role in determining the semimajor axis values of ultra-short-period planets and our findings do not exclude tidal migration. 

\subsection{Where Do Planets Form?}\label{where}

As previously mentioned, theoretical studies show that resonant chain systems of small planets are assembled through convergent migration---a planetary inward migration process. This suggests the planet formation process can be viewed as one of two scenarios (although some combination of both could be operating):
\begin{itemize}
    \item Disk-driven assembly: The planet-forming zone is located at a semimajor axis value that is greater than the outermost planet in the resonant chain and the chain is assembled by subsequent disk-driven migration.
    \item Modified in-situ: Planets are initially formed near the dust sublimation zone with disk-driven migration operating to produce the innermost planets (inside $R_{dust}$) followed by subsequent  planet formation that forms planets at successively larger semimajor axis values in conjunction with disk-driven assembly of the resonant chain.
\end{itemize}

Unfortunately, the relatively small number of resonant chain systems hampers our ability to make broader inferences. However, previous work suggests that most multi-planet systems have a resonant or near-resonant configuration shortly after formation \citep{dai2024,hu2025,yi-xian2025} that is disrupted over time by dynamical effects. Using the multi-planet systems to augment the resonant systems (see Figure~\ref{resonant}, top panel), we can localize the planet-forming zones for the two scenarios above. If multi-planet systems form via disk-driven assembly, the efficient planet forming zone is $\gtrsim 100 R_{PMS}$ for solar type stars (PMS frame) and $\gtrsim 100 R_{\star}$ for stars (MS frame) ranging from $0.05-2.0M_{\odot}$, consistent with theoretical models that form planets primarily at a few au \cite[cf][]{mordasini2009}. If multi-planet systems form via a modified in-situ process, the planet forming zone covers a region $\sim15 - 100 R_{\star}$. 

Locating the planet formation zone $\gtrsim 100 R_{\star}$ (corresponding to $\gtrsim \: 0.5$ au for a Solar analog)  has several interesting consequences. Models that invoke planet formation at the silicate condensation radius \citep[e.g.,][]{flock2019} or the disk dead zone inner boundary \citep[DZIB; e.g.,][]{chatterjee2014} are not forbidden, but there are questions about how they would be integrated into a picture where planets are forming at $\gtrsim100 R_{\star}$ and then undergo disk-driven migration to $\sim 10 R_{\star}$. Scenarios of in-situ formation in the inner disk may need to be modified by the addition of a subsequent stage involving a modest (of order unity) level of migration of the planets or where the sequential locations of planet formation are preferentially selected to be near low-order mean motion resonances.

It may be possible to distinguish observationally between the disk-driven assembly and modified in-situ formation scenarios for multi-planet systems. A key prediction of in situ formation models in which planets form inside the water ice line is that their bulk compositions are relatively volatile poor \citep[e.g.,][]{chatterjee2014}. Such characteristics have been inferred from an analysis of the close-in Super-Earth/Sub-Neptune population by \citet{2021MNRAS.503.1526R}. Note, in this scenario all close-in Super-Earth/Sub-Neptunes form with rocky/iron-rich cores of a few Earth masses and then accrete a primordial H/He atmosphere that is a few percent by mass of the planet. The Super-Earth and Sub-Neptune populations, separated by the radius valley, are then created later via photoevaporation of the innermost planet atmospheres. Thus the extent to which planetary cores are observed to be volatile-poor and the shape of the radius valley appears to be sculpted by photoevaporation is evidence in favor of in situ formation models.

Another potential test of formation scenarios is via the composition of planetary primordial atmospheres. multi-planet systems assembled via the disk-driven scenario should have similar C/O ratios in their atmospheres because they will have formed beyond the carbon soot line \citep{bergin2023} and inside of the CO and CO$_{2}$ ice lines \citep{oberg2011}. In contrast, multi-planet systems that are produced by modified in-situ formation are likely to have planets that have been formed at locations both interior and exterior of the carbon soot line. Thus, absent processes that produce substantial modifications to the C or O inventory, we would expect the planets in multi-planet systems to have similar C/O values if the system was produced by disk-driven assembly and systematic differences between the C/O values for interior and exterior planets if the system was produced through a modified in-situ process. 

\subsection{Single vs multi-planet Systems}

Our definition of single-planet systems is based on observational results and it is possible that additional undiscovered planets exist. A KS test comparing the full set of single planets to multi-planet system planets finds that these are separate populations (p-value $=10^{-60}$). It is evident in Figure~\ref{results} that single planet and multi-planet systems have different distributions in the semimajor axis vs $R_{\star}$ space with an obvious, vertically extended cluster of large radius single planets centered around $\sim 7 R_{\star}$, $1.5R_{\odot}$. This cluster of single planets raises the question of whether single and multi-planet systems have the same formation locations and migration histories. 

\begin{figure}[t]
\hspace{-0.25in} \\
\centering
\includegraphics[scale=1.0, angle=0]{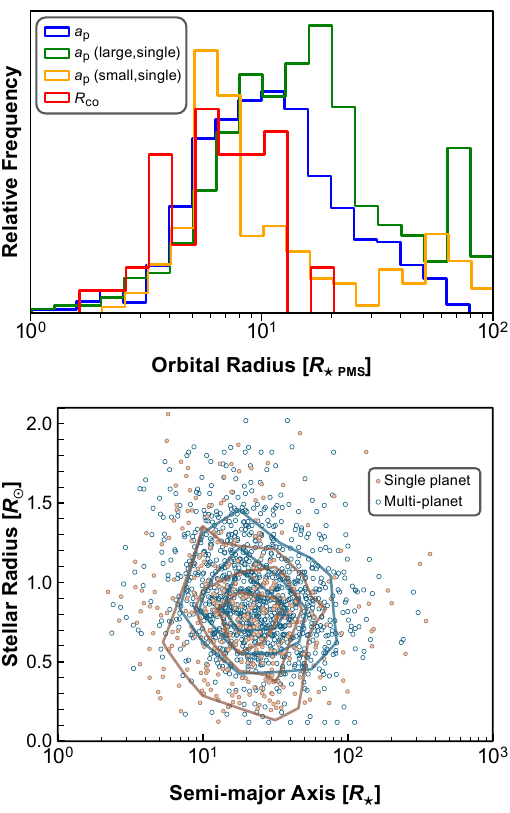}\\
\caption{{\it (a) Top:} the observation bias corrected semimajor axis in single planet systems is consistent with migration being halted at the disk co-rotation radius. {\it (b) Bottom:} contour plot for the single-planet and multi-planet systems with planet density contours set at 1,2,3 $\sigma$ shows that when planets with $R_{p}>10R_{\oplus}$ are removed, the single planet and multi-planet populations are similar, although not identical.} 
\label{single_multi}
\end{figure}

The majority of planets in the multi-planet systems are not in the gas giant category, while numerous single planet hot Jupiter type systems are known. Hot Jupiters may be produced by high eccentricity migration followed by tidal circularization \citep{dawson2018}, which is a fundamentally different process than the combined disk-driven followed by dynamical migration scenario that we argue is highly plausible for the multi-planet systems. If we divide the single planet systems into two categories, one with $R_{p}<10R_{\oplus}$ and $R_{p}>10R_{\oplus}$ (see Figure~\ref{single_multi} top panel), we find that while both categories stop in the region of $R_{co}$, the small planets stop at smaller semimajor axis values; previous theoretical work has suggested that migration is halted for different locations for larger and smaller planets \citep{zhuoya2026,soto2026}. If we filter the single planet systems to remove planets with $R_{p}>10R_{\oplus}$, we find that single and multi-planet system distributions become more self-similar (see Figure~\ref{single_multi}) but are likely still different distributions based on a KS test, which shows that the single planet systems with $R_{p}<10R_{\oplus}$ are different distributions (p-value = 0.0004). Perhaps this difference in distribution does not imply a difference in formation, as previous work indicates that single planet and multi-planet systems have a similar origin \citep{weiss2018b}.

\subsection{Comparison with Radius Gap \& Habitable Zone} \label{radiusgap}

Two concepts widely discussed in the context of exoplanets are the radius gap \citep{owen2013,fulton2017} and the habitable zone \citep{kasting1993}. Displaying these concepts in the context of our planet sample and analytical framework has the potential to reveal interesting trends. 

\begin{figure}[t]
\hspace{-0.25in} \\
\centering
\includegraphics[scale=1.0, angle=0]{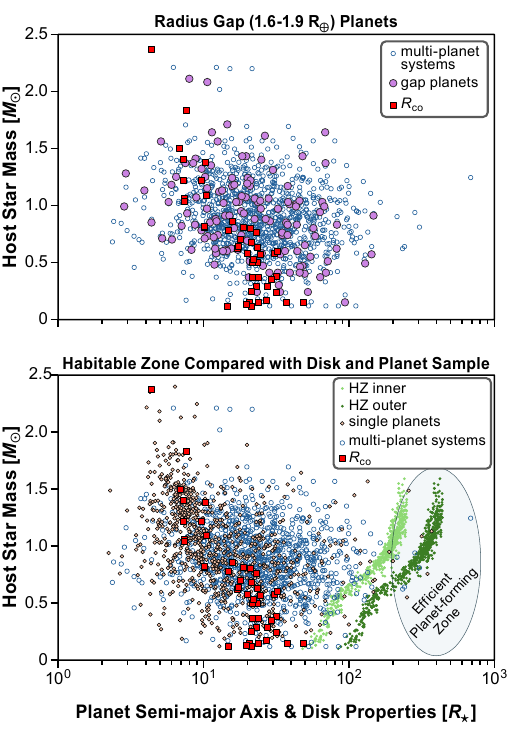}\\
\caption{{\it (a) Top:} Planets in the radius gap do not show a clear increase in semimajor axis value as function of $R_{\star}$. {\it (b) Bottom:} Assuming the multi-planet systems in our sample are assembled by disk-driven migration, the habitable zone lies inside of, or intersects with, the zone of efficient planet formation.} 
\label{radius_gap}
\end{figure}

The reduction in occurrence rate for planets with $\approx 1.6 \leq R_{p} \leq 1.9 R_{\oplus}$ has been attributed to several different potential physical origins including photoevaporation \citep{owen2013,owen2018}, core powered mass loss \citep{ginzburg2018}, and disk-temperature modulated gas accretion \citep{lee2022}; the models for all of these processes are consistent with the observations. Recent work \citep{dai2024} shows the probability of a planet being part of a resonant chain drops significantly for planets in the radius gap, raising the possibility that dynamical effects, such as impacts, may contribute to envelope loss. When we plot the semimajor axis values of members of the multi-planet systems with $1.6 \leq R_{p} \leq 1.9 R_{\oplus}$, there is no clear trend of increasing semimajor axis values of the radius gap planets with $R_{\star}$ as might be expected if photoevaporation was the dominant mechanism for envelope loss or disk-temperature-modulated gas accretion (see Figure~\ref{radius_gap}(a)).  The absence of any clear trend could be consistent with impacts playing a role, as suggested by \cite{dai2024}. A more detailed comparison of the gap planets semimajor axis distribution with specific model predictions is beyond the scope of this work.

To compare the habitable zone with the multi-planet semimajor axis and disk properties, radial distributions with respect to the host star, we have elected to use the ``conservative'' habitable zone defined as a $0.95-1.67$ au annulus around a Solar twin \citep{kopparapu2013} based on the range of planet temperatures where liquid water could exist on the planet's surface. As Figure~\ref{radius_gap}(b) shows, the habitable zone can be co-located with one of the two potential planet formation zones identified in this study. Even when the habitable zone lies inside the region of efficient planet formation, if resonant and multi-planet systems are formed through convergent migration, planets are likely to be initially common in the habitable zone for stars with $\leq M_{\odot}$. However, as Figure~\ref{sma_histo} shows, subsequent dynamical migration removes a large fraction of habitable zone planets for solar-type stars. Conversely, for low mass stars, dynamical migration can increase the occurrence rate of planets in the habitable zone. In this case, the time scale for dynamical migration matters. Concerns have been raised about how the flare history of low mass stars impacts habitability \citep[cf][and references there in]{estrela2020}. If dynamical migration operates on timescales of $\gtrsim 100$ My, flare activity in the host star may have materially decreased prior to the dynamically-driven arrival of additional habitable zone planets. A better understanding of dynamical migration may be important for projects such as the Habitable Worlds Observatory when assembling the sample of stars to be observed.


\subsection{Future Work}


The large difference in our study between the numbers of known planets in multi-planet systems and the number of measurements of protoplanetary disk properties is an area that needs to be addressed in the future. There are hints in the sample of 47 targets we use from the ODYSSEUS survey that the relationship between $R_{i}$ and $R_{co}$ may depend on stellar mass. For example, $R_{i}$ and $R_{co}$ appear to be substantially closer for M dwarf host stars. Ideally, the measurements of disk properties also need to be conducted using uniform methods. As noted in \S {\ref{migration}}, additional observations of disk properties would fill in mass ranges that need better sampling, which would benefit the comparison of disk and planet semimajor axis values. Also, when we have a larger sample of young planets, as mentioned in \S {\ref{dynamic}}, we will be able to better constrain when disk migration occurs. 

There are some prospects for dramatic improvement in the measurement of protoplanetary disk properties such as $R_{i}$ and $R_{co}$ through the proposed EVE mission \citep{macgregor2025}, and NASA SMEX concept, which will measure the properties of hundreds of disks. Targeting young clusters, EVE would provide wide-field, simultaneous, multiband photometery in the near-UV, visible, and near-infrared bands. The EVE measurements would enable the  discovery of young planets, the characterization of stellar flare energy distributions, and hundreds of protostellar disk $R_{i}$ and $R_{co}$ measurements. The ability to densely sample the $R_{i}$ and $R_{co}$ distributions would allow high fidelity comparison of planet location and disk properties and would also provide the measurements needed to see if there are systematic differences in protoplanetary disks as a function of stellar mass.

\section{Conclusions}

Comparing the properties of protoplanetary disks to the semimajor axis of multi-planet systems provides constraints on where planets form and how the semimajor axis evolves. Our study finds the following:
\begin{itemize}
    \item The stellar radius is a natural length scale for identifying trends in disk properties and semimajor axis values for transiting multi-planet systems. Whether the most useful stellar radius is the PMS or main sequence value depends on the parameters under consideration. 
    \item Planet occurrence rate falls dramatically inside the corotation radius, suggesting that disk torques near the corotation radius may act to halt inward planet migration. 
    \item Disk-driven migration occurs frequently and is needed to explain the location of inner planets in multi-planet systems. 
    \item If multi-planet systems are initially assembled in resonant chains, then subsequent dynamical migration profoundly sculpts the planet semimajor axis distribution from $\approxeq 20 - 300$ $R_{\star}$.
    \item If resonant systems are a common outcome of multi-planet system formation and are assembled via convergent migration, a zone of efficient planet formation exists $>100R_{\star}$.
    \item Pure in-situ planet formation is uncommon but modified in-situ planet formation incorporating modest migration and an outwardly moving zone of planet formation from $\approx 15-100 R_{\star}$ could be consistent with the data.
\end{itemize}

A clear area for future work is enlarging the survey data for disk properties. In our study, the exoplanets outnumber the disk property measurements by about 34:1. Dense sampling of $R_{i}$ and $R_{co}$ for a range of stellar masses from $\approx 0.2-2 M_{\odot}$ would better determine the range of values these can take on and allow a better understanding of the hints we see in the current data of a stellar-mass-dependent relationship between $R_{i}$ and $R_{co}$. Our study demonstrates that statistically understanding properties of the inner disk allows population level inference on where planets form and how much their orbits evolve. This is an area where the proposed EVE mission could make a significant impact.

\section*{Acknowledgments}
We thank Doug Lin for helpful discussions of processes for halting planet migration. 
This research has made use of the NASA Exoplanet Archive, which is operated by the California Institute of Technology, under contract with the National Aeronautics and Space Administration under the Exoplanet Exploration Program. The research was carried out at the Jet Propulsion Laboratory, California Institute of Technology, under a contract with the National Aeronautics and Space Administration (80NM0018D0004).

\clearpage
\bibliography{ref}
\end{document}